\documentclass[12pt]{iopart}
\usepackage{graphicx}
\newcommand{\V}{\mathcal{V}}
\newcommand{\vev}[1]{\left\langle #1 \right\rangle}
\newcommand{\Sf}{S_{\mathrm{free}}}
\newcommand{\Si}{S_{\mathrm{int}}}
\newcommand{\ep}{\epsilon}
\newcommand{\Om}{\mathcal{O}_m}
\newcommand{\Ol}{\mathcal{O}_\lambda}
\newcommand{\N}{\mathcal{N}}
\newcommand{\phitwo}{\left[\frac{1}{2} \phi^2 \right]_{\mathrm{MS}}}
\newcommand{\phifour}{\left[\frac{1}{4!} \phi^4\right]_{\mathrm{MS}}}
\newcommand{\dphitwo}{\left[\frac{1}{2} (\partial_\mu \phi)^2
 \right]_{\mathrm{MS}}} 
\newcommand{\dtwophitwo}{\left[\phi
    \partial^2 \phi\right]_{\mathrm{MS}}}
\begin{document}
\title[RG vs ERG]{On the relation between RG and ERG}
\author{H Sonoda}
\address{Physics Department, Kobe University, Kobe 657-8501, Japan}
\ead{hsonoda@kobe-u.ac.jp}

\begin{abstract}
We discuss how the ordinary renormalization group (RG) equations arise
in the context of Wilson's exact renormalization group (ERG) as
formulated by Polchinski.  We consider the $\phi^4$ theory in four
dimensional euclidean space as an example, and introduce a particular
scheme of parameterizing the solutions of the ERG equations.  By
analyzing the scalar composite operators of dimension two and four, we
show that the parameters obey mass independent RG equations.  We
conjecture the equivalence of our parameterization scheme with the MS
scheme for dimensional regularization.
\end{abstract}
\submitto{\JPA}
\maketitle

\section{Introduction}

The exact renormalization group (ERG) was introduced by K.~G.~Wilson
as a proper language to define continuum limits in quantum field
theory.  \cite{wk} A key ingredient is the theory space $\mathcal{S}$.
Given a set of fields, it consists of all possible theories (i.e.,
lagrangians or actions) with the same cutoff (or defined on the same
lattice).  A renormalization group (RG) transformation acts on
$\mathcal{S}$, and it consists of two steps:
\begin{enumerate}
\item integrating out high momentum modes
\item rescaling space to restore the same cutoff
\end{enumerate}
We incorporate short-distance physics into the action, leaving
long-distance physics for further integration of field variables.
Starting from a theory, RG transformations generate a flow of theories
along which the same physics is kept.  Only the physical momentum
scale of the cutoff becomes smaller along the flow.  In this setup,
the continuum limits form a finite dimensional subspace $\mathcal{S}
(\infty)$ of $\mathcal{S}$.  It is centered around a fixed point, and
its dimension is given by the number of relevant (or renormalized)
parameters.  The continuum limits can be obtained as the long distance
limit of theories finely tuned to criticality.  Using physical units,
this prescription defines the limit of the infinite UV cutoff.  RG
transformations of the renormalized parameters are obtained simply by
restricting ERG on $\mathcal{S} (\infty)$.

It was Polchinski who first introduced ERG into perturbative field
theory.\cite{joe} Polchinski's ERG gives a concrete realization of
Wilson's ERG, but it differs from Wilson's in two aspects:
\begin{enumerate}
\item no rescaling of space
\item artificial splitting of the action into the free and interacting
  parts
\end{enumerate}
The first point is not essential since we can easily modify
Polchinski's ERG to incorporate rescaling.  The second point is an
unavoidable nature of perturbation theory.

In fact a more serious difference lies elsewhere.  To be concrete, let
us consider $\phi^4$ theory in four dimensional euclidean space.  The
(perturbatively) renormalized theory has two parameters, which we can
take as a squared mass $m^2$ and a coupling $\lambda$.  Hence,
$\mathcal{S} (\infty)$ is two dimensional, and therefore we need only
one parameter to distinguish various ERG flows.
\begin{figure}[h]
\centerline{\includegraphics{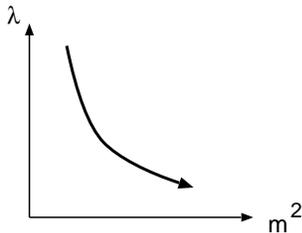}}
\caption{Wilson's ERG: only one parameter to specify an ERG flow}
\end{figure}
But this is not the case in Polchinski's ERG.  In
sect.~\ref{asymptotic} we will show that the action satisfying
Polchinski's ERG differential equation depends on three parameters
$(m^2, \lambda; \mu)$:
\begin{equation}
S (\Lambda; m^2, \lambda; \mu)
\end{equation}
where $\Lambda$ is the cutoff, decreasing along each ERG flow.  The
momentum scale $\mu$ is introduced as the scale where $m^2, \lambda$
are defined.  The two flows
\begin{equation}
(m^2, \lambda; \mu)\quad\&\quad \left( m^2 \e^{2t}, \lambda; \mu \e^t \right)
\end{equation}
correspond to different choices of mass units, and they are trivially
equivalent.  Let us denote the beta function of $\lambda$ as $\beta
(\lambda)$, and the anomalous dimension of $m^2$ as $\beta_m (\lambda)$.
Then, for an infinitesimal $\Delta t$, the flow
\begin{equation}
\left( m^2 \left(1 + \Delta t \beta_m (\lambda) \right),\, \lambda
\left( 1 + \Delta t \beta (\lambda) \right); \mu \e^{- \Delta t}
\right)
\end{equation}
is physically equivalent to the flow $(m^2, \lambda; \mu)$, but the
problem is that the two solutions
\begin{equation}
S (\Lambda; m^2, \lambda; \mu)\quad\&\quad S (\Lambda; m^2 (1 +
\Delta t \beta_m), \lambda + \Delta t \beta; \mu (1-\Delta t))
\end{equation}
do not overlap.  It is easy to see why.  The ERG differential equation
depends on $m^2$ explicitly, and the two solutions solve two different
ERG differential equations.
\begin{figure}[h]
\centerline{\includegraphics{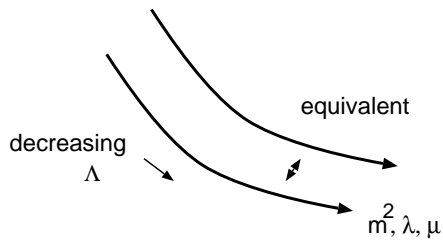}}
\caption{Polchinski's ERG: three parameters to specify an ERG flow}
\end{figure}
It is this redundancy of parameters which characterizes Polchinski's
ERG.  This makes the derivation of $\beta, \beta_m$ not
straightforward for Polchinski's ERG.

The purpose of this paper is twofold.  First we show how a momentum
scale $\mu$ enters the solution of Polchinski's ERG equations.  This
is done in sects.~2, 3, 4.  Second we derive $\beta$, $\beta_m$, and
the anomalous dimension $\gamma$ of the scalar field in the framework
of Polchinski's ERG.  Our derivation relies on the technique of
composite operators.  By studying the $\mu$ dependence of the action,
we will show how $\beta, \beta_m, \gamma$ arise naturally from ERG.
All this is done in sects.~5, 6, 7.

Not long after Polchinski's work, Hughes and Liu have looked at the
relation between ERG and RG.\cite{hl} Besides some technicalities, the
main difference from the present work is that they have overlooked the
difference between Wilson's and Polchinski's ERG.  This neglect makes
their results valid only at the lowest non-trivial orders in
perturbation theory.

In an unpublished work \cite{beta} we have obtained the same results
for $\beta, \beta_m, \gamma$ using Polchinski's ERG.  The present work
is based upon an entirely different technique of composite operators
which, the author hopes, makes the paper easier to follow.

\section{Polchinski's formulation of Wilson's exact renormalization
  group}

We consider a real scalar field theory in four dimensional euclidean
space.  Let $S$ be the action so that the correlation functions are
given in the momentum space as
\begin{eqnarray}
&&\vev{\phi (p_1) \cdots \phi (p_{2n})}_S \cdot (2\pi)^4 \delta^{(4)}
  (p_1 + \cdots + p_{2n}) \nonumber\\ &&\quad\equiv \left( \int
  [d\phi] \,\rme^S \right)^{-1} \int [d\phi] \phi (p_1) \cdots \phi
  (p_{2n}) \,\rme^{S}
\end{eqnarray}
$S$ consists of a free part $\Sf$ and an interaction part $\Si$:
\begin{equation}
S = \Sf + \Si
\end{equation}
where
\begin{eqnarray}
\Sf &\equiv& - \frac{1}{2} \int_p \phi (p) \phi (-p) \frac{p^2 +
  m^2}{K(p/\Lambda)}\qquad \left(\int_p \equiv \int \frac{d^4 p}{(2
  \pi)^4}\right) \\ \Si &\equiv& \sum_{n=1}^\infty \frac{1}{(2n)!}
  \int_{p_1,\cdots,p_{2n}} \phi (p_1) \cdots \phi (p_{2n})\nonumber\\
  &&\qquad \times (2\pi)^4 \delta^{(4)} (p_1+\cdots+p_{2n}) \cdot
  \V_{2n} (\Lambda; p_1, \cdots, p_{2n})
\end{eqnarray}
We take the cutoff function $K(x)$ as a decreasing positive function
of $x^2$ with the properties that $K(x) = 1$ for $0 \le x^2 \le 1$,
and that it decreases sufficiently fast for large $x^2 \gg 1$.
(See the left in Fig.~3.)
\begin{figure}[h]
\centerline{\includegraphics{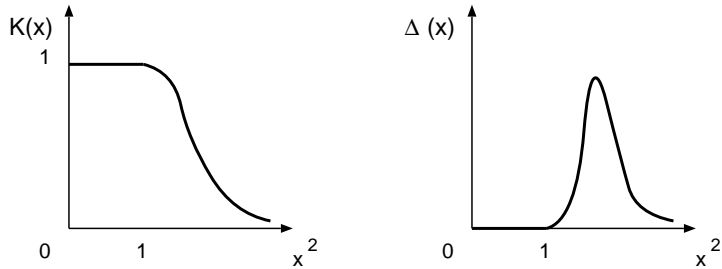}}
\caption{The behavior of the cutoff function $K$ and its derivative
  $\Delta$}
\end{figure}
$\Lambda$ is the momentum cutoff, suppressing the field fluctuations
of momenta larger than $\Lambda$.  For convenience, we introduce 
\begin{equation}
\Delta (x) \equiv - 2 x^2 \frac{d K(x)}{d x^2}
\end{equation}
This vanishes for $0 \le x^2 < 1$, and is positive for $x^2 > 1$.
(See the right in Fig. 3.)

Given an action $S$, the correlation functions can be computed
perturbatively using 
\begin{equation}
\frac{K(p/\Lambda)}{p^2 + m^2}
\end{equation}
as the propagator, and $\V_{2n}$ as the interaction vertices.  Thanks
to the rapid decrease of $K (p/\Lambda)$ for $p^2 \gg \Lambda^2$, the
Feynman integrals are free from ultraviolet divergences.  The use of a
finite ultraviolet cutoff $\Lambda$ is compensated, however, by the
presence of an infinite number of interaction terms.  Each interaction
vertex $\V_{2n}$ results from the integration over field momenta
larger than $\Lambda$, and it is local in the sense that it can be
expanded in powers of $m^2$ and external momenta if they are small
compared to appropriate powers of $\Lambda$.  In the extreme limit
$\Lambda \to 0$, $\V_{2n}$ reduces to the connected $2n$-point
correlation function.

As we change $\Lambda$, we wish to change $S(\Lambda)$ so that physics
is preserved.  This requirement translates into Polchinski's
differential equation \cite{joe}
\begin{eqnarray}
- \Lambda \frac{\partial}{\partial \Lambda} S (\Lambda) &=& \int_p
\frac{\Delta (p/\Lambda)}{p^2 + m^2} \Bigg[ \frac{p^2 +
    m^2}{K(p/\Lambda)} \phi (p) \frac{\delta S (\Lambda)}{\delta \phi
    (p)} \nonumber\\ && \qquad + \frac{1}{2} \left( \frac{\delta S
    (\Lambda)}{\delta \phi (p)} \frac{\delta S (\Lambda)}{\delta \phi
    (-p)} + \frac{\delta^2 S (\Lambda)}{\delta \phi (p) \delta \phi
    (-p)} \right) \Bigg] \label{pol}
\end{eqnarray}
This equation assures that the correlation functions, calculated with
$S(\Lambda)$, are independent of $\Lambda$.  To be more precise, the
following combinations do not depend on $\Lambda$:
\begin{equation}
\frac{1 - \left( K (p/\Lambda)\right)^{-1}}{p^2 + m^2} + \vev{\phi (p)
 \phi (-p)}_{S(\Lambda)} \frac{1}{K(p/\Lambda)^2}
\end{equation}
for the two-point, and
\begin{equation}
 \vev{\phi (p_1) \cdots \phi (p_{2n})}_{S(\Lambda)} \prod_{i=1}^{2n}
 \frac{1}{K (p_i/\Lambda)}
\end{equation}
for the higher-point functions $n > 1$.  Despite the use of a finite
ultraviolet cutoff, the action can contain physics of momentum scales
up to the highest $\Lambda$ for which $S$ is well defined.

\section{Solution by initial conditions}

A standard way of solving (\ref{pol}) is by imposing a set of initial
conditions at a large cutoff $\Lambda_0$.  In \cite{joe} Polchinski
showed the possibility of choosing initial conditions such that $S
(\Lambda)$, for any finite $\Lambda$, has a limit as $\Lambda_0 \to
\infty$.  (This was shown, of course, within perturbation theory.)
For example, we can adopt the following mass independent form:
\begin{eqnarray}
&&\V_2 (\Lambda_0; p,-p)\nonumber\\
&&\quad= \Lambda_0^2 \,g\left(\lambda, \ln (\Lambda_0/\mu)\right)
+ m^2 z_m \left(\lambda, \ln (\Lambda_0/\mu)\right) + p^2 z \left(\lambda, \ln
(\Lambda_0/\mu)\right)\\
&&\V_4 (\Lambda_0; p_1,\cdots,p_4) = - \lambda \left\lbrace 1 + z_\lambda
(\lambda, \ln (\Lambda_0/\mu)) \right\rbrace\\
&&\V_{2n \ge 6} (\Lambda_0; p_1,\cdots,p_{2n}) = 0
\end{eqnarray}
$g, z_m, z$, and $z_\lambda$ are all given as power series in
$\lambda$ whose coefficients depend logarithmically on $\Lambda_0$.
Here $\mu$ is an arbitrary finite momentum scale.  It is not only
necessary to render the argument of the logarithm dimensionless, but
it also acquires an important physical interpretation as the scale
where the renormalized parameters $\lambda, m^2$ are defined.  

To understand better how $\mu$ enters the continuum limit, let us look
at the one-loop two-point vertex.  Solving (\ref{pol}) we obtain
\begin{eqnarray}
&&\V_2 (\Lambda; p,-p) - \V_2 (\Lambda_0; p,-p)\nonumber\\
&& =\frac{-\lambda}{4} \left(\Lambda_0^2 - \Lambda^2\right) \int_q
\frac{\Delta (q)}{q^2} + \frac{\lambda}{(4 \pi)^2} m^2 \ln
(\Lambda_0/\Lambda) \nonumber\\
&& - \frac{\lambda}{2} \int_\Lambda^{\Lambda_0} \frac{d
  \Lambda'}{\Lambda'} \int_q \Delta (q/\Lambda') \frac{m^4}{q^4 (q^2 +
  m^2)}
\end{eqnarray}
where the last double integral on the right-hand side is finite as
$\Lambda_0 \to \infty$.  Therefore, for $\V_2 (\Lambda; p,-p)$ to have
a finite limit as $\Lambda_0 \to \infty$, we can take
\begin{eqnarray}
g (\lambda, \ln (\Lambda_0/\mu)) &=& \frac{\lambda}{4} \int_q
\frac{\Delta (q)}{q^2}\\
z_m (\lambda, \ln (\Lambda_0/\mu)) &=& - \frac{\lambda}{(4 \pi)^2} \ln
(\Lambda_0/\mu)\\
z (\lambda, \ln (\Lambda_0/\mu)) &=& 0
\end{eqnarray}
so that
\begin{eqnarray}
\V_2 (\Lambda; p,-p) &=& \frac{\lambda}{4} \Lambda^2 \int_q \frac{\Delta
  (q)}{q^2} - \frac{\lambda}{(4 \pi)^2} m^2 \ln (\Lambda/\mu)
\nonumber\\
&& - \frac{\lambda}{2} \int_\Lambda^{\infty} \frac{d \Lambda'}{\Lambda'}
\int_q \Delta (q/\Lambda') \frac{m^4}{q^4 (q^2 + m^2)} 
\label{twopoint-initial}
\end{eqnarray}
in the continuum limit.  Note that $\V_2 (\Lambda_0; p,-p)$ cannot
depend on $\Lambda$ since it gives an initial condition to the
differential equation (\ref{pol}) which determines the $\Lambda$
dependence of $\V_2 (\Lambda; p,-p)$.  We are thus obliged to introduce
$\mu$ to make sense of the logarithm of $\Lambda_0$ for $z_m$.
Similarly, the choice
\begin{equation}
z_\lambda (\lambda, \ln (\Lambda_0/\mu)) = \frac{3 \lambda}{(4 \pi)^2}
\ln (\Lambda_0/\mu)
\end{equation}
makes the continuum limit of $\V_4 (\Lambda)$ finite.  Arbitrary
finite constants can be added to $z_m$, $z_\lambda$, $z$, amounting to
further finite renormalization.

The dependence of the theory on an arbitrary momentum scale $\mu$ is
familiar from the standard renormalization theory.  In the above we
have adopted a minimal scheme in which the expansions of $z_m,
z_\lambda, z$ have at least one power of $\ln (\Lambda_0/\mu)$, with
no part independent of $\ln (\Lambda_0/\mu)$.  Giving a different
value to $\mu$ amounts to finite renormalization.  This permits us to
interpret $\mu$ as the scale where the renormalized parameters $m^2,
\lambda$ are defined.  Unless we choose to parameterize $S (\Lambda)$
in terms of physical parameters such as a physical squared mass and a
physical coupling, the introduction of $\mu$ is inevitable.

\section{Solution by asymptotic conditions \label{asymptotic}}

There is another way of solving (\ref{pol}), which is more convenient
for relating ERG to the RG of renormalized parameters.

We first note that perturbative renormalizability amounts to
\begin{equation}
\V_{2n} (\Lambda; p_1, \cdots, p_{2n}) \stackrel{\Lambda \to
  \infty}{\longrightarrow} 0
\end{equation}
for $2 n \ge 6$.  Besides the squared mass $m^2$ that appears in the
ERG equation (\ref{pol}) itself, the solutions depend on a coupling
parameter $\lambda$.  Hence, we can write the vertices in the form
\begin{eqnarray}
&&\V_{2n} (\Lambda; p_1, \cdots, p_{2n})
  \nonumber\\ && \quad = \Lambda^{y_{2n}} v_{2n} \left( \ln (\Lambda/\mu);
  \,p_1/\Lambda, \cdots, p_{2n}/\Lambda; \,m^2/\Lambda^2, \lambda\right)
\end{eqnarray}
where
\begin{equation}
y_{2n} \equiv 4 - 2n
\end{equation}
and $\mu$ is a momentum scale introduced to characterize the asymptotic
behavior of the vertices.  Expanding in powers of $m^2/\Lambda^2$ and
$p^2/\Lambda^2$ we obtain
\begin{eqnarray}
&&\V_2 (\Lambda; p, -p) \nonumber\\
&& \quad= \Lambda^2 a_2 \left(\ln
(\Lambda/\mu); \lambda \right)  + m^2 b_2 \left(\ln
(\Lambda/\mu); \lambda \right) + p^2 c_2 \left(\ln
(\Lambda/\mu); \lambda \right) + \cdots
\end{eqnarray}
Similarly, we obtain
\begin{equation}
\V_4 (\Lambda; p_1, \cdots, p_4) = a_4 \left(\ln
(\Lambda/\mu); \lambda \right) + \cdots
\end{equation}
The parts represented by dots are proportional to inverse powers of
$\Lambda$, and vanish in the limit $\Lambda \to \infty$.

At each order of $\lambda$, the asymptotic parts of the vertices are
given as finite degree polynomials of $\ln (\Lambda/\mu)$.  The
coefficients of a polynomial, say $P(\ln (\Lambda/\mu))$, depend on the
choice of $\mu$.  For example, the coefficient of the constant term,
$P(0)$, can be made to vanish by choosing a particular value for
$\mu$.

The ERG differential equation specifies only the $\Lambda$ dependence
of the vertices.  Hence, the $\Lambda$ independent parts of $b_2, c_2,
a_4$ do not get fixed uniquely.  One way of removing the ambiguity is
to adopt the following conditions \cite{ms}:
\begin{eqnarray}
&& b_2 (0; \lambda) = c_2 (0; \lambda) = 0 \label{bc}\\
&& a_4 (0; \lambda) = - \lambda \label{a}
\end{eqnarray}
We call this MS (minimal subtraction), since it resembles the MS
scheme for dimensional regularization.  The $1/\ep$ in dimensional
regularization corresponds to $\ln (\Lambda/\mu)$.  (\ref{bc}, \ref{a})
are the obvious analogues of the absence of finite parts in $\ep$ in
the three renormalization constants.

Adopting MS, straightforward one-loop calculations give the following
results:
\begin{eqnarray}
&&\V_2 (\Lambda; p,-p) = \frac{\lambda}{4} \Lambda^2 \int \frac{\Delta
  (q)}{q^2} - \frac{\lambda}{(4\pi)^2} m^2 \ln (\Lambda/\mu) \nonumber\\
&&\qquad\qquad\qquad  - \frac{\lambda}{2}
  (m^2)^2 \int_\Lambda^\infty \frac{d\Lambda'}{{\Lambda'}^3} \int_q
  \frac{\Delta (q)}{q^4 (q^2 + m^2/{\Lambda'}^2)} \label{vtwo}
\end{eqnarray}
for the two-point vertex (this is the same as
(\ref{twopoint-initial})), corresponding to
\begin{equation}
b_2 (\ln (\Lambda/\mu); \lambda) = - \frac{\lambda}{(4\pi)^2} \ln
(\Lambda/\mu),\quad c_2 = 0
\end{equation}
and 
\begin{eqnarray}
&&\V_4 (\Lambda; p_1,\cdots,p_4) = - \lambda - \lambda \sum_{i=1}^4 \frac{1 -
    K(p_i/\Lambda)}{p_i^2 + m^2} \, \V_2 (\Lambda;
    p_i,-p_i)\nonumber\\
&& \,+ \frac{\lambda^2}{2} \int_q \left[ \frac{1 -
    K(q/\Lambda)}{q^2+m^2} \frac{1 - K\left( (q+p_1+p_2)/\Lambda
    \right)}{(q+p_1+p_2)^2 + m^2} - \frac{\left(1 -
    K(q/\Lambda)\right)^2}{q^4} \right]\nonumber\\
&& \,+ (\textrm{t, u-channels}) - 3 \frac{\lambda^2}{(4 \pi)^2} \ln
    (\Lambda/\mu) \label{vfour}
\end{eqnarray}
for the four-point vertex, corresponding to
\begin{equation}
a_4 (\ln (\Lambda/\mu); \lambda) = - \lambda \left( 1 + \frac{3
\lambda}{(4 \pi)^2} \ln (\Lambda/\mu) \right)
\end{equation}

In our MS scheme the action depends on the four parameters $m^2$,
$\lambda$, $\mu$, $\Lambda$, and we may denote the action as
\begin{equation}
S (\Lambda; m^2, \lambda; \mu)
\end{equation}
Using the cutoff function, we can define the renormalized correlation
functions
\begin{eqnarray}
&&\vev{\phi (p) \phi (-p)}_{m^2, \lambda; \mu} \equiv \frac{1 - \left(
K (p/\Lambda)\right)^{-1}}{p^2 + m^2} + \vev{\phi (p) \phi
(-p)}_{S(\Lambda)} \frac{1}{K(p/\Lambda)^2}\\ &&\vev{\phi (p_1) \cdots
\phi (p_{2n})}_{m^2, \lambda; \mu} \equiv \vev{\phi (p_1) \cdots \phi
(p_{2n})}_{S(\Lambda)} \prod_{i=1}^{2n} \frac{1}{K (p_i/\Lambda)}
\end{eqnarray}
which are independent of $\Lambda$.  Let us recall that in the MS
scheme for dimensional regularization, the $\mu$ dependence of the the
correlation functions is canceled by compensating $\mu$ dependence of
$m^2, \lambda$, and field normalization.  This is how the beta
functions of $m^2, \lambda$ are derived.  In the remainder of this
paper we wish to do the same for the MS solutions of Polchinski's
equation.

This is an appropriate place to comment on the work of Hughes and Liu
\cite{hl}.  They parameterize the action using three parameters:
\begin{eqnarray}
\rho_1 &\equiv& \V_2 (\Lambda; 0,0)\\
\rho_2 &\equiv& \frac{\partial}{\partial p^2} \V_2 (\Lambda; p,
-p)\Big|_{p=0}\\
\rho_3 &\equiv& \V_4 (\Lambda; 0,0,0,0)
\end{eqnarray}
A solution to (\ref{pol}) is specified by the conditions at $\Lambda
= \Lambda_R$:
\begin{equation}
\rho_1 = \rho_2 = 0,\quad \rho_3 = - \lambda
\end{equation}
$\Lambda_R$ obviously corresponds to our $\mu$.  There are two
problems.  One, which is crucial, is that the dependence of the action
on the choice of $\Lambda_R$ is neglected.  The other is that this is
not a mass independent scheme; the three parameters depend on $m^2$
non-trivially.  Approximate mass independence is obtained by taking
$\Lambda^2$ very large compared to $m^2$.  From the $\Lambda$
dependence of $\rho_{1,2,3}$, they have derived the beta function and
anomalous dimensions at the lowest non-trivial orders.  We will make a
further remark in sect.~\ref{simple}.

\section{Composite operators}

To derive the $\mu$ dependence of the action, we will use the
technique of composite operators.  (See, for example, \cite{becchi}.)
This section contains a brief summary.

Let $\Phi (p)$ be a composite operator of momentum $p$:
\begin{eqnarray}
\Phi (p) &=& \sum_{n=1}^{\infty} \frac{1}{(2n)!}
\int_{p_1,\cdots,p_{2n}} \phi (p_1) \cdots \phi (p_{2n}) \nonumber\\
&& \quad \times (2\pi)^4 \delta^{(4)} (p_1+\cdots+p_{2n}-p)\cdot
\Phi_{2n} (\Lambda; p_1,\cdots,p_{2n})
\end{eqnarray}
The $\Lambda$ dependence
\begin{eqnarray}
- \Lambda \frac{\partial}{\partial \Lambda} \Phi (p)
&=& \int_q \frac{\Delta (q/\Lambda)}{q^2 + m^2} \Bigg( \frac{q^2 +
  m^2}{K(q/\Lambda)} \phi (q) \frac{\delta}{\delta \phi (q)} \nonumber\\
&& +  \frac{\delta S}{\delta \phi (-q)} \frac{\delta}{\delta \phi (q)} 
 + \frac{1}{2} \frac{\delta^2}{\delta \phi (q) \delta \phi
   (-q)} \Bigg) \Phi (p)
\end{eqnarray}
guarantees that the correlation functions
\begin{equation}
\vev{\Phi (p) \phi (p_1) \cdots \phi (p_{2n})}_{m^2, \lambda; \mu}
\equiv \vev{\Phi (p) \phi (p_1) \cdots \phi (p_{2n})}_{S(\Lambda)}
\prod_{i=1}^{2n} \frac{1}{K(p_i/\Lambda)}
\end{equation}
are independent of $\Lambda$.  Composite operators of dimension $d$
satisfy the asymptotic conditions
\begin{equation}
\Phi_{2n > d} (\Lambda; p_1, \cdots, p_{2n})
\longrightarrow 0 \quad \textrm{as}\quad\Lambda \to \infty
\end{equation}

\subsection{Composite operators in the MS scheme}

For concreteness and also for later convenience, we consider scalar
composite operators of dimension $2$ and $4$.  We will show how to
construct composite operators using an analogous MS condition.

A dimension $2$ operator $\Phi$ has the asymptotic behavior:
\begin{equation}
\Phi_2 (\Lambda; p_1, p_2) = a_2 ( \ln (\Lambda/\mu) ) + \cdots
\end{equation}
where we suppress the $\lambda$ dependence of $a_2$, and the dots
denote the part vanishing in the limit $\Lambda \to \infty$.  In the
MS scheme we define $\Phi$ by imposing the asymptotic condition:
\begin{equation}
a_2 (0) = 1
\end{equation}
We will denote this operator as
\begin{equation}
\phitwo (p)
\end{equation}

Next we consider an operator $\Phi$ of dimension $4$.  It must satisfy
the following asymptotic behavior:
\begin{eqnarray}
&& \Phi_2 (\Lambda; p_1, p_2) = \Lambda^2 a_2 (\ln (\Lambda/\mu)) + m^2
  b_2 (\ln (\Lambda/\mu)) \nonumber\\ &&\qquad\qquad\qquad - (p_1 p_2)
  c_2 (\ln (\Lambda/\mu)) - (p_1^2 + p_2^2) c'_2 (\ln (\Lambda/\mu)) +
  \cdots\\ && \Phi_4 (\Lambda; p_1, \cdots, p_4) = a_4 (\ln
  (\Lambda/\mu)) + \cdots 
\end{eqnarray}
In the MS scheme we can define three linearly independent operators as
follows:
\begin{itemize}
\item $\phifour (p)$ satisfying
\begin{equation}
a_4 (0) = 1,\quad c_2 (0) = c'_2 (0) = b_2 (0) = 0
\end{equation}
\item $\dphitwo (p)$ satisfying
\begin{equation}
c_2 (0) = 1,\quad a_4 (0) = c'_2 (0) = b_2 (0) = 0
\end{equation}
\item $\dtwophitwo (p)$ satisfying
\begin{equation}
c'_2 (0) = 1,\quad a_4 (0) = c_2 (0) = b_2 (0) = 0
\end{equation}
\end{itemize}
For zero momentum we have only two linearly independent operators since
\begin{equation}
\dphitwo (0) = - \frac{1}{2} \dtwophitwo (0)
\end{equation}
Hence, an arbitrary dimension $4$ composite operator with zero
momentum is given as
\begin{equation}
\Phi = x \,m^2 \phitwo + y \phifour + z \dphitwo 
\end{equation}
where the $\Lambda$-independent coefficients can be extracted from the
asymptotic behavior:
\begin{equation}
x = b_2 (0),\quad y = a_4 (0),\quad z = (c_2 - 2 c'_2) (0)
\end{equation}

\section{Composite operators in terms of the action $S$}

The scalar composite operators of dimension $2, 4$ with zero momentum
are special in that they can be constructed directly out of the action
$S$.  It is easy to see why.  $\phitwo$ is the mass term, and it can
be obtained essentially by differentiating $S$ with respect to $m^2$.
$\phifour$ is the interaction term, and is obtained by differentiating
$S$ with respect to $\lambda$.  The hard part is to construct
$\dphitwo$; we need to use the equation of motion.

We examine three cases one by one.  For later convenience we introduce
the following expansions in $m^2$:
\begin{eqnarray}
  &&\V_4 (\Lambda; p, -p, 0, 0) = A_4 (\ln (\Lambda/\mu); p/\Lambda) +
  \frac{m^2}{\Lambda^2}  B_4 (\ln  (\Lambda/\mu); p/\Lambda) +
  \cdots\label{v4}\\ &&\V_6 (\Lambda; 
  p,-p,0,\cdots,0) = \frac{1}{\Lambda^2} A_6 (\ln (\Lambda/\mu); p/\Lambda)
  + \cdots \label{v6}
\end{eqnarray}
where $p$ is considered of order $\Lambda$. 

\subsection{$\Om$}

We define
\begin{eqnarray}
&&\Om \equiv - \frac{\partial S}{\partial m^2} \nonumber\\ &&\quad -
\int_q \frac{K(q/\Lambda) \left(1 - K(q/\Lambda)\right)}{\left(q^2 +
m^2\right)^2} \frac{1}{2} \left( \frac{\delta S}{\delta \phi (q)}
\frac{\delta S}{\delta \phi (-q)} + \frac{\delta^2 S}{\delta \phi (q)
\delta \phi (-q)} \right)
\end{eqnarray}
From the asymptotic behavior, we obtain
\begin{equation}
\Om = x_m \phitwo
\end{equation}
where
\begin{equation}
x_m \equiv 1 - \frac{1}{2} \int_q \frac{K(q)(1-K(q))}{q^4} A_4 (0; q)
\end{equation}
In the appendix we derive
\begin{equation}
\vev{\Om \phi (p_1) \cdots \phi (p_{2n})}_{m^2, \lambda; \mu}
 = - \frac{\partial}{\partial m^2} \vev{\phi (p_1) \cdots \phi
 (p_{2n})}_{m^2, \lambda; \mu} \label{om}
\end{equation}

\subsection{$\Ol$}

We define
\begin{equation}
\Ol \equiv - \frac{\partial S}{\partial \lambda}
\end{equation}
From the definition, it is straightforward to show
\begin{equation}
\vev{\Ol \phi (p_1) \cdots \phi (p_{2n})}_{m^2, \lambda; \mu}
 = - \frac{\partial}{\partial \lambda} \vev{\phi (p_1) \cdots \phi
 (p_{2n})}_{m^2, \lambda; \mu} \label{ol}
\end{equation}
From the asymptotic behavior, we obtain
\begin{equation}
\Ol = \phifour
\end{equation}

\subsection{$\N$}

We define
\begin{eqnarray}
&&\N \equiv - \int_q \phi (q) \frac{\delta S}{\delta \phi (q)}\nonumber\\
&& \quad - \int_q
\frac{K(q/\Lambda) \left(1 - K(q/\Lambda)\right)}{q^2 +
  m^2} \left( \frac{\delta S}{\delta \phi (q)} \frac{\delta
  S}{\delta \phi (-q)} + \frac{\delta^2 S}{\delta \phi (q) \delta \phi
  (-q)} \right)
\end{eqnarray}
We show, in the appendix, that
\begin{equation}
\vev{\N \phi (p_1) \cdots \phi (p_{2n})}_{m^2, \lambda; \mu}
= 2n \vev{\phi (p_1) \cdots \phi (p_{2n})}_{m^2, \lambda; \mu} \label{n}
\end{equation}
By examining the asymptotic behavior we obtain
\begin{equation}
\N = x_\N \,m^2 \phitwo + y_\N \phifour + z_\N \dphitwo
\end{equation}
where 
\begin{eqnarray}
x_\N &\equiv& 2 + \int_q K(q)(1-K(q)) \left( - \frac{B_4
  (0;q)}{q^2} + \frac{A_4 (0;q)}{q^4} \right)\\
y_\N &\equiv& - 4 \lambda - \int_q \frac{K(q)(1-K(q))}{q^2} A_6 (0;
q) \\
z_\N &\equiv& 2 - \int_q \frac{K(q)(1-K(q))}{q^2} C_4 (0;q)\label{zn}
\end{eqnarray}
In (\ref{zn}) $C_4$ is defined by
\begin{equation}
\frac{1}{\Lambda^2} \delta_{\mu\nu} C_4 (\ln (\Lambda/\mu); p/\Lambda)
\equiv \frac{1}{2} \frac{\partial^2}{\partial q_\mu \partial q_\nu}
\V_4 (\Lambda; p,-p,q,-q) \Big|_{m^2=q^2=0}
\end{equation}
where the angular average over $p_\mu$ is taken on the right-hand
side.  Using the results for $\Om$ and $\Ol$, we can rewrite this as
\begin{equation}
\N = \frac{x_\N}{x_m} m^2 \Om + y_\N \Ol + z_\N \dphitwo
\end{equation}

To conclude this section, we have shown that the three scalar
operators of dimension $2, 4$ can be constructed from $S$ as
\begin{eqnarray}
\phitwo &=& \frac{1}{x_m} \Om\\
\phifour &=& \Ol\\
\dphitwo &=& \frac{1}{z_\N} \left( \N - \frac{x_\N}{x_m} m^2 \Om -
y_\N \Ol \right)
\end{eqnarray}

\section{Beta function and anomalous dimensions}

With all the necessary tools in our hands, we are ready to derive the
ordinary RG equations for the renormalized correlation functions.
This is done by considering the $\mu$ dependence of the action:
\begin{equation}
\Psi \equiv \mu \frac{\partial}{\partial \mu} S(\Lambda; m^2, \lambda; \mu)
\end{equation}
Differentiating Polchinski's equation with respect to $\mu$, we obtain
\begin{eqnarray}
- \Lambda \frac{\partial}{\partial \Lambda} \Psi &=& \int_q
 \frac{\Delta (q/\Lambda)}{q^2 + m^2} \Bigg[ \frac{q^2 +
 m^2}{K(q/\Lambda)} \phi (q) \frac{\delta}{\delta \phi (q)}
 \nonumber\\ &&+ \frac{\delta S}{\delta \phi (-q)}
 \frac{\delta}{\delta \phi (q)} + \frac{1}{2} \frac{\delta^2}{\delta
 \phi (q) \delta \phi (-q)} \Bigg]\, \Psi
\end{eqnarray}
Hence, $\Psi$ is a composite operator of zero momentum.  This has
dimension $4$, and hence must be a linear combination of three
independent operators $\Om, \Ol, \N$.

We first expand $\Psi$ in terms of MS operators by examining its
asymptotic behaviors.  We find
\begin{eqnarray}
&&\Psi_2 (\Lambda; p,-p) = \Lambda^2 \dot{a}_2 (\ln (\Lambda/\mu);
  \lambda)\nonumber\\
&&\qquad\qquad\qquad  + m^2
\dot{b}_2 (\ln (\Lambda/\mu); \lambda) + p^2 \dot{c}_2 (\ln (\Lambda/\mu);
  \lambda) + \cdots\\
&&\Psi_4 (\Lambda; p_1,\cdots,p_4) =  \dot{a}_4 (\ln (\Lambda/\mu); \lambda) +
\cdots
\end{eqnarray}
where 
\begin{equation}
\dot{b}_2 (\ln (\Lambda/\mu); \lambda) \equiv \mu
\frac{\partial}{\partial \mu} b_2 (\ln (\Lambda/\mu); \lambda) = -
\Lambda \frac{\partial}{\partial \Lambda} b_2 (\ln (\Lambda/\mu);
\lambda), \cdots
\end{equation}
are the derivatives of the coefficients introduced in
sect.~\ref{asymptotic}.  Hence, $\Psi$ can be expanded as
\begin{equation}
\Psi = \dot{b}_2 (\lambda) m^2 \phitwo + \dot{c}_2 (\lambda) \dphitwo +
\dot{a}_4 (\lambda) \phifour
\end{equation}
where
\begin{equation}
\dot{b}_2 (\lambda) \equiv \dot{b}_2 (0; \lambda),\quad
\dot{c}_2 (\lambda) \equiv \dot{c}_2 (0; \lambda),\quad
\dot{a}_4 (\lambda) \equiv \dot{a}_4 (0; \lambda)
\end{equation}

Using the results of the previous section, we can rewrite the above
using $\Om, \Ol$, and $\N$ instead:
\begin{eqnarray}
\Psi &=& \frac{\dot{b}_2}{x_m} m^2 \Om + \frac{\dot{c}_2}{z_\N} \left(
\N - \frac{x_\N}{x_m} m^2 \Om - y_\N \Ol \right) + \dot{a}_4 \Ol\\
&=& \frac{1}{x_m} \left( \dot{b}_2 - \dot{c}_2 \frac{x_\N}{z_\N}
\right) m^2 \Om + \left( \dot{a}_4 - \dot{c}_2 \frac{y_\N}{z_\N}
\right) \Ol + \frac{1}{z_\N} \dot{c}_2 \N
\end{eqnarray}
We now define
\begin{eqnarray}
\beta (\lambda) &\equiv& - \left( \dot{a}_4 - \dot{c}_2
\frac{y_\N}{z_\N} \right)\\ \beta_m (\lambda) &\equiv& - \frac{1}{x_m}
\left( \dot{b}_2 - \dot{c}_2 \frac{x_\N}{z_\N} \right)\\ \gamma
(\lambda) &\equiv& - \frac{1}{z_N} \dot{c}_2
\end{eqnarray}
so that
\begin{equation}
\Psi \equiv \mu \frac{\partial S}{\partial \mu} = - \left( \beta_m m^2
\Om + \beta \Ol + \gamma \N \right) \label{main}
\end{equation}

It is now trivial to derive the $\mu$ dependence of the correlation
functions:
\begin{eqnarray}
 &&\mu \frac{\partial}{\partial \mu} \vev{\phi (p_1) \cdots \phi
 (p_{2n})}_{m^2, \lambda; \mu} \nonumber\\
&& = \vev{\mu \frac{\partial S}{\partial
 \mu} \phi (p_1) \cdots \phi (p_{2n})}_{m^2, \lambda; \mu}\nonumber\\
 && = \vev{\left( - m^2 \beta_m \Om - \beta \Ol - \gamma \N \right)
 \phi (p_1) \cdots \phi (p_{2n})}_{m^2, \lambda; \mu}\nonumber\\ && =
 \left( m^2 \beta_m \frac{\partial}{\partial m^2} + \beta
 \frac{\partial}{\partial \lambda} - 2 n \gamma \right) \vev{\phi
 (p_1) \cdots \phi (p_{2n})}_{m^2, \lambda; \mu}
\end{eqnarray}
where we have used (\ref{om},\ref{ol},\ref{n}).  This is nothing but
the usual RG equations for the renormalized correlation functions.
Hence, $\beta$ is the beta function of $\lambda$, and $\beta_m, \gamma
$ are the anomalous dimensions of $m^2$, $\phi$, respectively.

At 1-loop, we find
\begin{equation}
\dot{b}_2 (\lambda) = \frac{\lambda}{(4 \pi)^2},\quad \dot{c}_2
(\lambda) = 0,\quad \dot{a}_4 (\lambda) = 3 \frac{\lambda^2}{(4
  \pi)^2}
\end{equation}
We also find, to leading order in $\lambda$,
\begin{equation}
x_m = 1,\quad x_\N = 2,\quad y_\N = - 4 \lambda,\quad z_\N = 2
\end{equation}
Hence,
\begin{equation}
\beta = - 3 \frac{\lambda^2}{(4 \pi)^2},\quad \beta_m = -
\frac{\lambda}{(4 \pi)^2},\quad \gamma = 0
\end{equation}
reproducing the familiar results.  For two-loop calculations, see
\cite{beta} (and \cite{hl} for $\gamma$).

\section{Simplification in the Wegner-Houghton limit \label{simple}}

In the Wegner-Houghton limit \cite{wh} we take
\begin{equation}
K(x) = \theta (1-x^2) = \left\lbrace\begin{array}{c@{\quad}l}
1 & (x^2 < 1)\\ 0 & (x^2 > 1)\end{array}\right.
\end{equation}
This limit is known to introduce non-locality to the theory.  For
example, the inverse Fourier transform of the high-momentum propagator
\begin{equation}
\int_p \e^{i p r} \frac{1 - \theta (\Lambda^2 - p^2)}{p^2 + m^2}
\end{equation}
behaves as $\sin (\Lambda r)/r^2$ for $\Lambda r \gg 1$.  Despite
this, this limit has its own merit of bringing nice simplification to
ERG.

In the Wegner-Houghton limit we obtain
\begin{equation}
K(p/\Lambda) \left( 1 - K(p/\Lambda) \right) = 0
\end{equation}
Hence, $\Om, \Ol, \N$ become simply
\begin{equation}
\Om = - \frac{\partial S}{\partial m^2},\quad
\Ol = - \frac{\partial S}{\partial \lambda},\quad
\N = - \int_p \phi (p) \frac{\delta S}{\delta \phi (p)}
\end{equation}
Therefore, (\ref{main}) gets simplified to
\begin{equation}
\mu \frac{\partial S}{\partial \mu} = \left( \beta_m m^2
\frac{\partial}{\partial m^2} + \beta \frac{\partial}{\partial
  \lambda} + \gamma \int_p \phi (p) \frac{\delta}{\delta \phi (p)}
\right) S
\end{equation}
meaning that a change of $\mu$ can be compensated by appropriate
changes in $m^2, \lambda$ and field normalization.  By examining the
asymptotic behavior of this equation, we obtain the following
equations:
\begin{eqnarray}
&& \left( \mu \frac{\partial}{\partial \mu} - \beta \partial_\lambda
  \right) \left(1 - b_2 (\ln (\Lambda/\mu); \lambda)\right) = \left(
  \beta_m + 2 \gamma \right) \cdot \left(1 - b_2 (\ln (\Lambda/\mu);
  \lambda)\right)\\ && \left(\mu \frac{\partial}{\partial \mu} -
  \beta\partial_\lambda \right) \left(1 - c_2 (\ln (\Lambda/\mu);
  \lambda)\right) = 2 \gamma \cdot \left(1 - c_2 (\ln (\Lambda/\mu);
  \lambda)\right)\\ && \left( \mu \frac{\partial}{\partial \mu} -
  \beta \partial_\lambda \right) a_4 (\ln (\Lambda/\mu); \lambda) = 4
  \gamma \cdot a_4 (\ln (\Lambda/\mu); \lambda)
\end{eqnarray}
We introduce a running coupling $\bar{\lambda} (t; \lambda)$ by
\begin{equation}
\frac{\partial}{\partial t} \bar{\lambda} (t; \lambda) = \beta \left(
\bar{\lambda} (t; \lambda) \right),\quad
\bar{\lambda} (0; \lambda) = \lambda
\end{equation}
so that the above equations are solved by
\begin{eqnarray}
1 - b_2 (\ln (\Lambda/\mu); \lambda) &=& \exp \left[ - \int_0^{\ln
    (\Lambda/\mu)} dt \,\left\lbrace \beta_m (\bar{\lambda} (-t; \lambda)) + 2
    \gamma (\bar{\lambda} (-t; \lambda)) \right\rbrace\right]\\
1 - c_2 (\ln (\Lambda/\mu); \lambda) &=&  \exp \left[ - 2 \int_0^{\ln
    (\Lambda/\mu)} dt \,\gamma (\bar{\lambda} (-t; \lambda)) \right]\\
a_4 (\ln (\Lambda/\mu); \lambda) &=& - \exp \left[ - 4 \int_0^{\ln
    (\Lambda/\mu)} dt \,\gamma (\bar{\lambda} (-t; \lambda))
    \right] \bar{\lambda} (-\ln (\Lambda/\mu); \lambda)
\end{eqnarray}
Hence, the asymptotic behavior of the vertices are expressed fully in
terms of $\beta, \beta_m, \gamma$.  Conversely, we can determine
$\beta, \beta_m, \gamma$ by calculating $b_2, c_2, a_4$.  At the
lowest non-trivial orders in $\lambda$, the above equations are valid
for any choice of $K$.  This explains why the correct results were
obtained in \cite{hl}.  (Their $\rho_{1,2,3}$ correspond to $b_2$,
$c_2$, $a_4$, respectively.)

The above results are so similar to the well known relation for the
renormalization constants in the MS scheme for dimensional
regularization. \cite{th} From this, we conjecture that $\beta,
\beta_m, \gamma$ in the Wegner-Houghton limit are the same to all
orders in $\lambda$ to those in the MS scheme for dimensional
regularization.  This is supported by the 2-loop calculations of
$\gamma, \beta_m$ which are scheme dependent. \cite{hl,beta} But we
have no other justification for this conjecture.

\section{Conclusions}

In this paper we have explained how the ordinary RG equations arise from
Polchinski's ERG equations.  We achieved this goal in two steps:
\begin{enumerate}
\item introduction of MS scheme --- We characterize the solutions to
  ERG differential equations by their asymptotic behavior at large
  cutoff.  A renormalization scale $\mu$ is introduced to organize the
  logarithmic dependence of the asymptotic vertices on the cutoff.
\item derivation of the $\mu$ dependence of the solution to ERG --- We
  have shown that the $\mu$ derivative of the action is a composite
  operator of zero momentum, and therefore it can be expanded by the
  three operators whose properties we understand very well.
\end{enumerate}
Throughout the paper we have emphasized the difference between
Wilson's ERG and Polchinski's ERG.  They share the same spirit, but
there are crucial differences that make the derivation of $\beta,
\beta_m, \gamma$ somewhat non-trivial for Polchinski's ERG.

Is there any way of modifying Polchinski's ERG to get something more
like Wilson's?  There is.  Instead of following the $\Lambda$
dependence of $S(\Lambda; m^2, \lambda; \mu)$, we follow the $\mu$
dependence of
\begin{equation}
\bar{S} (\mu; m^2, \lambda) \equiv S(\mu; m^2, \lambda; \mu)
\end{equation}
This $\bar{S}$ is characterized by the large $\mu$ behaviors
\begin{eqnarray}
&&\bar{\V}_2 (\mu; m^2, \lambda; p,-p) = \mu^2 a_2
(\lambda) + \cdots\\
&&\bar{\V}_4 (\mu; m^2, \lambda; p_1, \cdots, p_4) = - \lambda +
\cdots\\
&&\bar{\V}_{2n \ge 6} (\mu; m^2, \lambda; p_1, \cdots, p_{2n}) =
\cdots
\end{eqnarray}
which are devoid of logarithms.  (Although the asymptotic parts have
no explicit $\mu$ dependence, the parts vanishing asymptotically
depend on $\mu$.  This can be seen from the two one-loop examples
(\ref{vtwo}, \ref{vfour}).)  Under an infinitesimal change of $\mu \to \mu
(1 - \Delta t)$, we also change $m^2, \lambda$ as
\begin{equation}
m^2 \longrightarrow m^2 \left(1 + \Delta t \cdot \beta_m \right),\quad
\lambda \longrightarrow \lambda + \Delta t \cdot \beta
\end{equation}
Then, the new ERG differential equation gives
\begin{eqnarray}
&&\Bigg( - \mu \frac{\partial}{\partial \mu} + \beta_m (\lambda) m^2
\frac{\partial}{\partial m^2} + \beta (\lambda)
\frac{\partial}{\partial \lambda} \nonumber\\
&&\qquad\qquad\qquad\qquad + \gamma (\lambda) \int_p \phi (p)
\frac{\delta}{\delta \phi (p)} \Bigg) \,\bar{S} (\mu; m^2, \lambda)
\end{eqnarray}
in terms of integrals over momenta of order $\mu$.  (This has been
done in \cite{beta}.)  By construction,
\begin{equation}
\bar{S} \left(\mu (1 - \Delta t); m^2 (1 + \Delta t \beta_m), \lambda +
\Delta t \beta \right)
\end{equation}
lies on the same ERG trajectory as $\bar{S} (\mu; m^2, \lambda)$.

As we have seen in this paper, the technique of composite operators is
a useful and perhaps essential tool for addressing formal questions
about perturbative ERG.  The composite operators have already been
shown to play essential roles in the applications of ERG to gauge
theories and non-linear sigma models.  (See for example,
\cite{becchi}.)  This is also the case in a forthcoming paper
\cite{qed} where we will apply the MS scheme of ERG to QED.

\appendix

\section{Derivation of $\Om$ and $\N$}

In this appendix we wish to derive the relation of the operators $\Om,
\Ol, \N$ to the action.  As preparation, we introduce two types of
operators.  (We call them operators though they are not necessarily
composite operators on their own.)

\subsection{Type 1 operator $\mathcal{O}_1 [f]$}

We define
\begin{equation}
\mathcal{O}_1 [f] \equiv \int_p f(p) \phi (p) \frac{\delta S}{\delta
  \phi (p)}
\end{equation}
where $f(p)$ is an arbitrary function of momentum.  This generates an
infinitesimal linear change of field:
\begin{equation}
\delta \phi (p) = - f (p) \phi (p)
\end{equation}
It is straightforward to derive the correlation functions of
$\mathcal{O}_1$:
\begin{eqnarray}
&&\vev{\mathcal{O}_1 [f] \phi (p_1) \cdots \phi (p_{2n})}_S \cdot (2
\pi)^4 \delta^{(4)} (p_1 + \cdots + p_{2n}) \nonumber\\ && = \int_p
f(p) \int [d\phi] \phi (p) \frac{\delta S}{\delta \phi (p)} \phi (p_1)
\cdots \phi (p_{2n}) \e^S\nonumber\\ && = \int_p f(p) \int [d\phi]
\phi (p) \phi (p_1) \cdots \phi (p_{2n}) \frac{\delta \e^S}{\delta
\phi (p)}\nonumber\\ && = - \int_p f(p) \int [d\phi] \phi (p)
\frac{\delta}{\delta \phi (p)} \left( \phi (p_1) \cdots \phi (p_{2n})
\right) \e^S\nonumber\\ && = - \sum_{i=1}^{2n} f(p_i) \cdot \vev{\phi
(p_1) \cdots \phi (p_{2n})}_S \cdot (2 \pi)^4 \delta^{(4)} (p_1 +
\cdots + p_{2n})
\end{eqnarray}
where we have neglected the derivative acting on $\phi (p)$, since it
does not contribute to the connected part of the correlation function.
Thus, we obtain
\begin{equation}
\vev{\mathcal{O}_1 [f] \phi (p_1) \cdots \phi (p_{2n})}_S = -
\sum_{i=1}^{2n} f(p_i) \cdot \vev{\phi (p_1) \cdots \phi (p_{2n})}_S
\end{equation}

\subsection{$\mathcal{O}_2 [C]$}

We define
\begin{equation}
\mathcal{O}_2 [C] \equiv \int_p C (p) \frac{1}{2} \left(
 \frac{\delta S}{\delta \phi (p)} \frac{\delta S}{\delta \phi (-p)} +
 \frac{\delta^2 S}{\delta \phi (p) \delta \phi (-p)} \right)
\end{equation}
where $C(p) = C(-p)$.  This generates an infinitesimal non-linear
change of field:
\begin{equation}
\delta \phi (p) = - C (p) \frac{\delta S}{\delta \phi (-p)}
\end{equation}
We can show
\begin{eqnarray}
\vev{\mathcal{O}_2 [C] \phi (p) \phi (-p)}_S &=& C (p)\\
\vev{\mathcal{O}_2 [C] \phi (p_1) \cdots \phi (p_{2n})}_S &=& 0\quad
(n > 1)
\end{eqnarray}
The first equation is derived as follows:
\begin{eqnarray}
&&\vev{\mathcal{O}_2 [C] \phi (p_1) \phi (p_2)}_S \cdot (2 \pi)^4
  \delta^{(4)} (p_1+p_2) \nonumber\\ && = \int_q
C(q) \frac{1}{2} \int [d\phi] \left( \frac{\delta S}{\delta \phi (q)}
\frac{\delta S}{\delta \phi (-q)} + \frac{\delta^2 S}{\delta \phi (q)
\delta \phi (-q)} \right) \phi (p_1) \phi (p_2) \e^S\nonumber\\ && =
\int_q C(q) \frac{1}{2} \int [d\phi] \phi (p_1) \phi (p_2) \frac{\delta^2
  \e^S}{\delta \phi (q) \delta \phi (-q)} \nonumber\\
&& = \int_q C(q) \frac{1}{2} \int [d\phi] \e^S \frac{\delta^2}{\delta \phi
  (q) \delta \phi (-q)} \phi (p_1) \phi (p_2)\nonumber\\
&& = C (p_1) (2\pi)^4 \delta^{(4)} (p_1 + p_2)
\end{eqnarray}
For $n > 1$, the connected part receives no contribution.

We note that the right-hand side of Polchinski's ERG equation
(\ref{pol}) is the sum of type 1\&2 operators.

\subsection{Derivation of $\Om$}

We first consider the two-point function.  Since
\begin{equation}
\vev{\phi (p) \phi (-p)}_{m^2,\lambda} \equiv \frac{1 -
  1/K(p/\Lambda)} {p^2 + m^2} + \vev{\phi (p) \phi (-p)}_S \cdot
  \frac{1}{K(p/\Lambda)^2} 
\end{equation}
we obtain
\begin{eqnarray}
&&- \frac{\partial}{\partial m^2} \vev{\phi (p) \phi
    (-p)}_{m^2,\lambda}\nonumber\\ && = \frac{1 -
    1/K(p/\Lambda)}{(p^2+m^2)^2} - \vev{\phi (p) \phi (-p)
    \frac{\partial S}{\partial m^2}}_S
    \frac{1}{K(p/\Lambda)^2}\nonumber\\ && = \left( -
    \frac{K(p/\Lambda)(1-K(p/\Lambda))}{(p^2+m^2)^2} - \vev{\phi (p)
    \phi (-p) \frac{\partial S}{\partial m^2}}_S \right)
    \frac{1}{K(p/\Lambda)^2}
\end{eqnarray}
For the higher-point functions, we simply obtain
\begin{eqnarray}
&&- \frac{\partial}{\partial m^2} \vev{\phi (p_1) \cdots \phi
  (p_{2n})}_{m^2,\lambda} \nonumber\\
&& \quad= - \vev{\phi (p_1) \cdots \phi (p_{2n})
  \frac{\partial S}{\partial m^2}}_S \cdot \prod_{i=1}^{2n}
  \frac{1}{K(p_i/\Lambda)} 
\end{eqnarray}
Thus, using a type 2 operator $\mathcal{O}_2 [C]$ for
\begin{equation}
C (p) \equiv \frac{K(p/\Lambda)(1-K(p/\Lambda))}{(p^2+m^2)}
\end{equation}
we obtain
\begin{eqnarray}
&&- \frac{\partial}{\partial m^2} \vev{\phi (p_1) \cdots \phi
  (p_{2n})}_{m^2,\lambda} \nonumber\\
&&\quad = \vev{\phi (p_1) \cdots \phi (p_{2n})
  \left( - \frac{\partial S}{\partial m^2} - \mathcal{O}_2 [C]
  \right)}_{m^2, \lambda}
\end{eqnarray}
for any $n = 1,2,\cdots$.  This defines $\Om$.

\subsection{Derivation of $\N$}

We consider the type 1 operator $\mathcal{O}_1 [f]$ corresponding to
$f = -1$:
\begin{equation}
\mathcal{O}_1 [f] = - \int_p \phi (p) \frac{\delta S}{\delta \phi (p)}
\end{equation}
Then, we obtain
\begin{equation}
\vev{\mathcal{O}_1 [f] \phi (p_1) \cdots \phi (p_{2n})}_S
= 2n \vev{ \phi (p_1) \cdots \phi (p_{2n})}_S
\end{equation}
for any $n$.  On the other hand, we have
\begin{eqnarray}
&&2 \vev{\phi (p) \phi (-p)}_{m^2,\lambda} = -
\frac{2(1-1/K(p/\Lambda))}{p^2+m^2} + 2 \vev{\phi (p) \phi (-p)}_S
\cdot \frac{1}{K(p/\Lambda)^2}\nonumber\\ && \quad = \left(
-\frac{2K(p/\Lambda)(1-K(p/\Lambda))}{p^2+m^2} + \vev{\phi (p) \phi
(-p) \mathcal{O}_1 [f]}_S \right) \frac{1}{K(p/\Lambda)^2}
\end{eqnarray}
and
\begin{eqnarray}
&&2 n \vev{\phi (p_1) \cdots \phi (p_{2n})}_{m^2,\lambda} = 2n
\vev{\phi (p_1) \cdots \phi (p_{2n})}_S \cdot \prod_{i=1}^{2n}
\frac{1}{K(p_i/\Lambda)} \nonumber\\ && \quad = \vev{\phi (p_1) \cdots
\phi (p_{2n}) \mathcal{O}_1 [f]}_S \cdot \prod_{i=1}^{2n}
\frac{1}{K(p_i/\Lambda)}
\end{eqnarray}
for $n > 1$.  Thus, using a type 2 operator $\mathcal{O}_2 [C]$ for
\begin{equation}
C (p) \equiv - \frac{2K(p/\Lambda)(1-K(p/\Lambda))}{p^2+m^2}
\end{equation}
we obtain
\begin{equation}
2 n \vev{\phi (p_1) \cdots \phi (p_{2n})}_{m^2,\lambda}
= \vev{\phi (p_1) \cdots \phi (p_{2n}) \left( \mathcal{O}_1 [f] +
  \mathcal{O}_2 [C] \right)}_{m^2, \lambda} 
\end{equation}
for any $n = 1, 2, \cdots$.  This defines $\N$.

\end{document}